# INVESTIGATION BY THE METHOD THE RAMAN OF SPECTROSCOPY OF ALLOCATION OF MOLECULES IN TERNARY MIX-CRYSTALS

## M. A. Korshunov

*LV Kirensky Institute of Physics, Siberian Branch of Russian Academy of Science, Krasnoyarsk, 660036 Russia*
*E-mail: mkor@iph.krasn.ru*

**Abstract:** The Method a Raman of spectroscopy studies allocation of molecules in ternary mix-crystals of a p-dibromobenzene of p-dichlorobenzene and p-bromochlorobenzene. It is shown, that the mutual concentration of builders depends on requirements of growing. Was possibly as a uniform modification of concentration of all builders along a specimen, and a wavy modification of concentration of two substances.

For a synthesis of large and homogeneous single crystals and fathoming of the mechanism of propagation the knowledge of regularity of allocation of impurity on a crystal depending on requirements of growing is necessary. Real crystals contain impurities. Besides in practice mix-crystals are widely used. Their properties depend on how impurity is proportioned by volume a single crystal. Better if allocation of builders of by volume mixed crystal uniform. That it to achieve the knowledge of regularity of allocation of impurities on a crystal is necessary. The knowledge of the mechanism of crystal growth also enables to influence it. In a number of operations [1] it is shown, that for two-component solid solutions depending on requirements of growing of single crystals on method Bridzhmena it is is possible both a uniform modification of concentration of builders, and wavy. In-process it speaks periodic emersion and a thinning away of blocks during propagation boundary lines which the segregation of impurity occurs. Is of interest to carry out analogous examinations for ternary mix-crystals and whether will become clear in similar crystals to be observed analogous appearances.

As physical properties of crystals are largely defined by dynamics of a lattice which is extremely sensitive to structural changes and composition of crystals that a method Raman of spectroscopy is rather convenient tool of diagnostic and optical monitoring [2].

p-dichlorobenzene, p-dibromobenzene and p-bromochlorobenzene have been chosen. These substances are isomorphous and crystallize in one space group $P2_1/a$ with two molecules in lattice cell and form among themselves two-component mix-crystals of substitution at any concentrations of builders [3]. Ternary single crystals pellucid, that allows to gain qualitative spectrums. Single crystals of solid solutions of studied substances have been grown and Raman spectra in the field of the lattice and intramolecular oscillations are gained (up to 400 $cm^{-1}$) earlier these spectrums were not studied. In figure 1 spectrums of intramolecular oscillations of ternary mix-crystals are resulted at concentration of a p-dibromobenzene 50 mol. %, p-dichlorobenzene 40 mol. % and p-bromochlorobenzene 10 mol. %. According to operation [3], the line in a spectrum of a p-dibromobenzene =212.0 $cm^{-1}$ (ag) matches to valence vibration C-Br, and a line with =327.0 $cm^{-1}$ (ag) in p-dichlorobenzene to valence vibration C-Cl an analogous line in p-bromochlorobenzene =261.0 $cm^{-1}$ (ag). Using a relationship intensity the symmetrical valence vibrations in mix-crystals para substitution benzene the modification of concentration of builders longwise single crystals is revealed. Examinations of different specimens it was spent at the same parameters of a data-acquisition equipment. In-process concentration of builders was measured in molar unities.

Spectrums of the lattice oscillations of studied ternary solid solutions are similar to spectrums of builders, it, apparently, confirms, that mix-crystals are formed as substitution. Single crystals have been grown on method Bridzhmena. The ampoule with substance was hauled down from hot in cold field with different velocities $V=8\times10^{-6}$ cm/s and above up to $15\times10^{-6}$ cm/s. The temperature gradient of the furnace was set by various winding of a spiral and made $dt/dl=7.6$ a grad/cm. At more low speed the monotonous modification of concentration of builders along an axis of a single crystal is observed. It was defined on a monotonous modification of intensity of lines of the symmetrical valence vibrations of molecules of p-dichlorobenzene, a p-dibromobenzene and p-bromochlorobenzene in Raman a spectrum longwise a specimen.

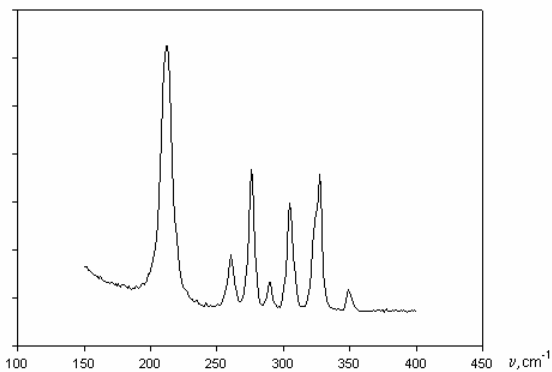

Figure 1. A spectrum of intramolecular oscillations of the ternary mixed crystal of a p-dibromobenzene - p-dichlorobenzene - p-bromochlorobenzene.

In figure 1 association of relative concentration of a p-dibromobenzene to p-dichlorobenzene (a) and is shown to p-bromochlorobenzene (b). Initial concentration of p-dibromobenzene $C^{Br}_0$ in a charge of the explored specimens made 50 mol. %. Apparently, in process of propagation of single crystals concentration of impurity increases.

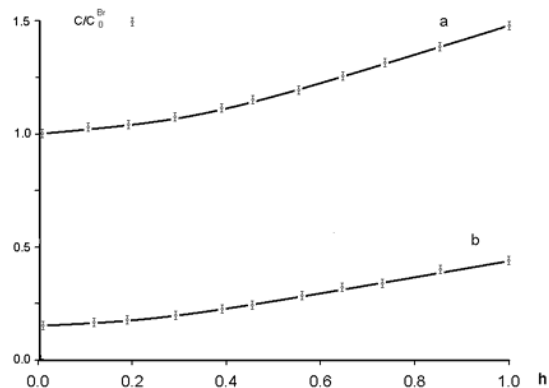

Figure 2. Allocation of impurity of p-dichlorobenzene (a) and p-bromochlorobenzene (b) longwise (h) a ternary crystal at low speed of propagation.

On fig. 3 graphs of allocation of a builder of p-dichlorobenzene $C/C^{Br}_0$ (a) on an axis of a single crystal in solid solutions are presented at greater growth rate concerning p-bromochlorobenzene (b). As we see they have wavy character thus magnification of concentration one will agree with decrease of concentration of another. Thus concentration of a p-dibromobenzene varies monotonically.

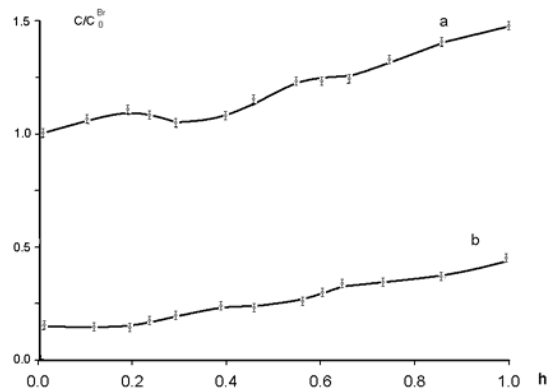

Figure 3. Allocation of impurity of p-dichlorobenzene (a) and p-bromochlorobenzene (b) longwise (h) trehkompanentnogo a crystal at a high speed of propagation.

As we see, changing requirements of growing it is possible to change allocation of molecules of impurity thus was possibly as a uniform modification of concentration of all builders along a specimen, and a wavy modification of concentration of two substances that will affect long-range and short-range order of a disposition of molecules. In too time other mutual allocation of molecules of impurity in three component crystals unlike two components is noted.

**Reference**.

[1]. Krivandina. **Kristallografya.23**, 2, (1978 372.
[2] M.A. Korshunov. Crystallography Report. Vol. **48**, No 3, (2003 525.
**[3]** Suzuki M., Ito M. Spectrochimica Acta. T. 25A.N 5.(1969) 1017.
**[4]** A. I. Kitaigorodskii, Rentgenostrukturnyi Analiz, (X-ray Analysis), Nauka, Moscow, 1973